\documentclass[12pt]{spieman}
 
\usepackage{amsmath,amsfonts,amssymb,aas_macros,lineno}
\usepackage{graphicx}
\usepackage{setspace}
\usepackage{tocloft}
\usepackage{bm}
\usepackage{orcidlink}

\title{Concept integral field unit spectrometer instrument for the next-generation mm-wave cosmological surveys}

\cftpagenumbersoff{figure}
\cftpagenumbersoff{table} 

\begin{document} 

\author[a,*]{Attila Kov\'{a}cs\orcidlink{https://orcid.org/0000-0001-8991-9088}}
\author[a]{Garrett K. Keating\orcidlink{https://orcid.org/0000-0002-3490-146X}}
\author[b,c,d]{Thomas R. Greve\orcidlink{https://orcid.org/0000-0002-2554-1837}}
\author[a]{Timothy Norton}
\affil[a]{Center for Astrophysics $|$ Harvard \& Smithsonian, 60 Garden St, Cambridge, MA 02138, U.S.A.}
\affil[b]{Cosmic Dawn Center (DAWN)}
\affil[c]{DTU-Space, Elektrovej, Building 328 , 2800, Kgs. Lyngby, Denmark}
\affil[d]{Dept. of Physics and Astronomy, University College London, Gower Street, London WC1E 6BT, United Kingdom}

\maketitle

\begin{abstract}
Past millimeter-wave galaxy surveys have primarily probed the brightest starburst galaxies only and suffered heavily from confusion. The interpretation of existing surveys has also been hindered by the lack of reliable redshift indicators for measuring distances for entire samples. Thanks to recent advances in mm-wave detector technologies we can now overcome these limitations, and conduct the first truly volumetric surveys of star-forming galaxies at mm-wavelengths approaching the $L_*$ luminosities of typical galaxies, with $\sim$1000 redshift slices spanning most of the Cosmic star-forming volume ($z$$\sim$1--12) with nearly uniform mass and luminosity selection. We describe an instrument concept capable of delivering such surveys with the technologies available today, which can be built and operated on a ground-based mm-wave facility in the near future. Such integral field unit spectrometers can resolve and identify redshifts for up to to 25,000 star-forming galaxies per year even when operated on a 10-m class telescope. On a larger aperture it can do the same faster or probe even deeper. We propose a collaboration open-source initiative to design, build, and operate one or several such cameras through the shared contributions of leading experts and telescopes from around the globe.
\end{abstract}


{\noindent \footnotesize\textbf{*}Attila Kov\'{a}cs,  \linkable{attila.kovacs@cfa.harvard.edu} }

\begin{spacing}{1}   

\section{Introduction}
\label{sec:intro}  

Stars form in dusty environments, obscured from direct view at the optical wavebands\cite{Devlin2009, Puget2006}. The optical/UV light from the young stellar populations is absorbed by the dust, and then re-emitted as cold ($\sim$35\,K) thermal radiation in the far-infrared (FIR) bands, peaking near $\lambda$$\sim$100\,$\mu$m in the rest frame\cite{Kovacs2010}. Even the Milky Way -- with modest star-formation rates of $\sim$1\,M$_{\odot}$/year -- radiates about half of its total bolometric luminosity in the FIR bands as a result, while galaxies with significantly higher star-formation rates, such as Arp220, emit over 99\% of the luminosities at FIR wavelengths, and are difficult to spot optically, especially at higher redshifts.

Cosmic star formation has peaked around redshift $z$$\sim$2--3\cite{Chapman2003,Madau2014}. The bulk of the stars we see today formed during intense starbursts in that era, when galaxies would have been at their most luminous phase (thanks to short-lived populations of massive OB stars), yet largely obscured at optical wavelengths. Catching galaxies during their intense episodes of star-formation requires FIR/(sub)mm surveys.

Due to a strongly negative $K$-correction on the Rayleigh-Jeans side of the thermal spectral energy distribution (SED), a star-forming galaxy will appear similarly bright at $\sim$1\,mm wavelength regardless of its distance at $z$$\sim$1--12 (Fig.~\ref{fig:selection}). This flat mass-luminosity selection\cite{Kovacs2010, Staguhn2014} allows us to conduct unbiased luminosity-limited surveys of star-formation in the mm-band that spans much of the Cosmic history in which galaxies formed and evolved. This effect has been exploited for a number of FIR galaxy surveys\cite{Weiss2009, Oliver2012, Simpson2019}. The Herschel Multi-tiered Extragalactic Survey (HerMES)\cite{Oliver2012} alone has detected on the order of a 100,000 FIR galaxies with its long-wavelength (200--500\,$\mu$m) SPIRE\cite{SPIRE} instrument.

However, the flat selection at mm-wavelength is a mixed blessing. Because every line of sight can see star-forming galaxies at all distances equally, the detections alone do not inform us about the redshift of these objects; and the fields quickly get crowded (confused) as extremely bright galaxies fill up every beam on sky well before one could hope to individually detect the fainter, more typical Milky-Way-like galaxy populations. Attempts to use photometry as a way to estimate redshifts for these galaxies have proven ineffective at best\cite{Cox2023}, while optical/NIR spectroscopic follow-up often relies on heavily biased cross-identifications\cite{Pope2006} and can target at most a tiny fraction of the galaxies detected by the continuum surveys overall.

In this paper, we demonstrate that a $\sim$100-pixel integral field unit (IFU) spectrometer, with $R\sim$200 resolution, can fill the focal planes of many existing (or future) 10-m class (or larger) mm-wave telescopes, and cover the entire 1\,mm atmospheric window with $\sim$20,000 detectors. The technologies required for such an instrument exist 'today' and could enable volumetric galaxy surveys, down to around L$_{*}$ luminosities efficiently, yielding up to 23,000 galaxies per survey year. In Section \ref{sec:future-surveys} we describe future volumetric surveys including the possibility of Line Intensity Mapping (LIM). In Section \ref{sec:enabling} we list some of the technologies that enable an IFU spectometer 'today'. In Section \ref{sec:concept} we describe what a typical realization for an existing 10-m class telescope may entail, followed by Section~\ref{sec:scoping} on how the designs may be adjusted / scoped for other telescopes, technologies, or resources. Finally, to build such an instrument for the Greenland Telescope (GLT), and to facilitate parallel development of similar instrument we suggest and open-source collaboration in Section~\ref{sec:implementation}.

\section{Future Surveys}
\label{sec:future-surveys}

One way to address the confusion is to conduct surveys with larger telescope apertures. Increasing of aperture diameter by a factor $X$ allows detecting $X^2$ more sources before the confusion limit is reached. And with SMG number counts $\partial N/ \partial S$$\propto S^{-3.2}$ typically for the (bright) submillimeter galaxy (SMG) populations\cite{Weiss2009}, the confusion flux limit (i.e.\ mass-luminosity selection threshold) will scale inversely as $S_{\rm min}\sim$$X^{-0.6}$. So larger apertures can help probe a little deeper, but they cannot effectively solve the redshift identification issue.

The only way to provide redshift identification for the SMG population without introducing extreme bias is through spectroscopy in the (sub)millimeter bands directly, targeting the principal cooling lines such as those of the bright CO rotational ladder in molecular gas, or else those in atomic or ionized species such as CI (492\,GHz and 809\,GHz), CII (158\,$\mu$m), or similar. Because galactic lines are broad (many hundreds of km/s typically), a moderate degree of spectral resolution ($R$$\sim$200--1000) is typically sufficient for the quasi-optimal detection of such transitions from spatially unresolved galaxies.

The CO rotational ladder offers perhaps the best tool for redshift identification. The lower-$J$ CO transitions are typically thermalized with a $S(\nu)$$\sim$$\nu^2$ spectrum on the Rayleigh-Jeans side of their pseudo-blackbody spectral energy distributions (SEDs). SMGs are known to exhibit quasi-thermalized emission into the moderately high-$J$ ($\sim$10) transitions\cite{Weiss2007,Boogaard2020}. The observed spacing of the CO ladder is $\Delta\nu$$\sim$115.3\,GHz $(1+z)^{-1}$ for a source at redshift $z$. Thus, the number of transitions detectable in a broad-band spectrometer scales as $N_{\rm CO}(z)$$\sim$$(1+z)$ at higher $z$, while the measurement noise from co-adding fluxes from the necessary channels increases as $\sqrt N_{\rm CO}$. As such, the detectability of redshift, through the combination of multiple CO transitions in band, benefits from a negative $K$-correction also with an effective emissivity index $\beta=0.5$ due to the multiplicity of available transitions. 

\begin{figure}[!hbt]
\centering
\includegraphics[width=0.95\textwidth]{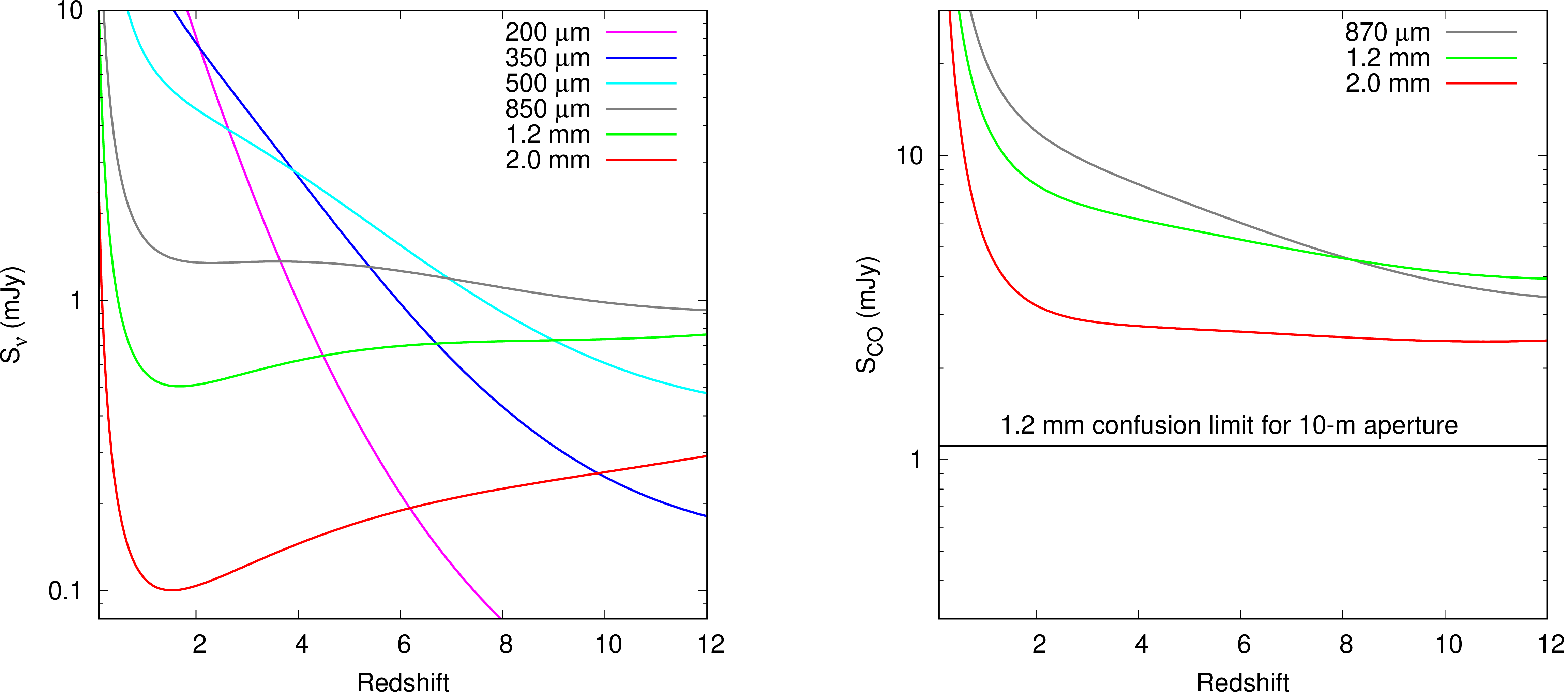}
\caption{\label{fig:selection} Typical redshift selection curves for a star-forming galaxies with $L$$\sim$10$^{12}$\,L$_{\odot}$ at various wavelengths. Left: continuum selection curves\cite{Kovacs2010}. Right: Estimated combined CO-selection curves assuming thermalized CO transitions. As the high-$J$ CO transitions may be sub-thermally excited, the actual selection curves will likely fall somewhat below this prediction at high $z$. The 1.2\,mm CO intensity confusion limit for an $R$$\sim$200 spectrometer on a 10-m class telescope is also shown, expected at around 3$\times$10$^{11}$\,L$_\odot$, which corresponds $\sim$3\,$L_*$ in the local Universe. However, this confusion limit is likely closer to, or below, $L_*$ for the distant Universe, given that global star-formation history peaked at $z$$\sim$2--3.}
\end{figure}

The ultimate result of which is that CO redshifts have a similarly flat mass-luminosity selection thresholds at millimeter wavelengths as the dust continuum (Fig.~\ref{fig:selection}). CO emission is also highly correlated with the thermal dust continuum in extragalactic star-forming environments\cite{Weiss2008}, both being equally effective tracers of the star-forming gas. As such, there is no fundamental need to detect SMGs by their far-infrared continuum emission separately, when detection via the CO ladder alone yields effectively the same mass-luminosity measure for star-forming sources. (Arguably, the CO selects star-formation more purely, since the mm-wave continuum may be contaminated by synchrotron emission, such as from active galactic nuclei [AGNs].)

\subsection{Volumetric Galaxy Surveys}

Because the CO ladder from a distant galaxy leaves a unique spectral imprint as a function of redshift, an $R$$\sim$200--1000 spectrometer covering octave-scale bandwidths can effectively sample $\sim$1000--5000 redshift slices above $z_{\rm min}$$\sim$230\,GHz$/ \Delta f  - 1$. Here $z_{\rm min}$ is the minimum redshift at which at least two CO transitions are detected in the observable bandwidth. Even if two galaxies overlap spatially, they will leave distinct spectral imprints so long as they are separated by $\Delta v_{\rm rest} > c / R$ in the rest-frame velocity space (or $dz/(1+z) > 1/R$ in redshift space). That alone reduces confusion dramatically when compared to continuum surveys with comparable beam sizes. As a result, an $R$$\sim$200--1000 integral field unit spectrometer instrument could detect up to 100--500 times more sources per beam than a continuum camera operated at the same wavelength before the spectral confusion limit is reached. Hence, an $R$$\sim$200--1000 survey can probe star-forming galaxies to a mass / luminosity threshold an order-of-magnitude below that of a continuum survey at the same wavelengths for the same-sized telescope. 

To put it into perspective, an $R$$\sim$200 spectroscopic survey at $\lambda$$\sim$1.2\,mm on a 10-m class telescope\cite{GLT-survey}, would be confused at a mass / luminosity threshold that is {\bf $\sim$20 times below} what Herschel/SPIRE could probe at its longest wavelength 500\,$\mu$m channel, while also peering further back in time. Consequently, it would resolve not only the most extreme ULIRG / HLIRG populations from before, but could for the first time systematically dip into the $L_*$ populations\cite{Cooray2005} (cf.\ Fig.~\ref{fig:selection} right panel) that dominate the makeup of galaxies overall -- while providing unambiguous redshift identification for every single detected source also. On a 30-m aperture it could reach about twice as deep still. Such surveys have been proposed for future 50-m class telescopes\cite{Kohno2018, Geach2019}, but not for the existing 10-m class telescopes, which could deliver significant results sooner.

\subsection{Line Intensity Mapping (LIM)}

A related application, for the same or very similar IFU spectrometer, is Line Intensity Mapping (LIM)\cite{Azadeh2022, GLT-LIM, Vallini2023}. In this case, the goal is not the detection and unique redshift identification of individually resolved galaxies, but rather the statistical (unresolved) detection of entire populations, and the statistical (e.g.\ 3D) correlations that characterize them. Because LIM targets galaxies with similar properties (e.g.\ velocity dispersion) as the above described survey, its requirements are inherently similar, specifically with $R$$\sim$200-1000 spectroscopic mapping\cite{Karkare2018, Karkare2022}. However, because the goal is not the individual detection of sources, LIM surveys probe much larger areas albeit at shallower depths.

While deep galaxy surveys that can be performed effectively in a single atmospheric window (e.g.\ at $\lambda$$\sim$1.2\,mm), LIM is best conducted over as much total bandwidth as possible, covering as many molecular and/or hyperfine transitions as possible. Thus, the ultimate LIM survey would require the next-level instrument upgrade, featuring simultaneous multi-band capability. An ideal mm-wave LIM instrument would cover the 3\,mm, 2\,mm, 1.2\,mm, and 850\,$\mu$m atmospheric windows with fully populated focal plane spectrometers simultaneously. As such it may contain up to four separate focal-planes, with a correspondingly higher total detector count to match. A multi-band implementation will inevitably require a far more complicated optical and cryogenic configuration than what a single-band IFU spectrometer can get away with. Sufficient stray-light and out-of-band light rejection will be significantly more challenging to achieve, and it requires more detectors to be read out. Nevertheless, a multiband IFU spectrometer is also possible with the right amount of ambition.

Both a single-band and multi-band spectroscopic instrument could satisfy both science goals (e.g. time-sharing to conduct two distinct surveys side-by-side). It is a matter of which type of science the instrument is designed to do more optimally. The resolved galaxy surveys and LIM are inherently complementary to one another, both in their survey modes (small-deep vs.\ large-shallow), and their scientific outputs (individually resolved galaxies vs.\ statistically identified unresolved populations). Ideally, we will be pursuing both science goals in the not too distant future.

\section{Enabling Technologies}
\label{sec:enabling}

Moderate resolution $R$$\sim$200--1000 spectroscopic surveys are a compelling way forward. They promise to overcome both limitations of past (sub)millimeter continuum surveys, with significantly lesser confusion {\em and} direct, unambiguous, in-band redshift identifications. To conduct such surveys, however, we would need multi-beam spectrometers with moderate spectral resolutions of $R$$\sim$200--1000 at mm-waves, preferably fully populating the field-of-view of a designated 10-m class or larger telescope. That in turn requires the following enabling technologies (scoped for a ground-based instrument, operated at $\lambda$$\sim$1\,mm, with $f/D$$\sim$3--10):

\begin{enumerate}
\item{Moderate resolution spectral channelizing in a footprint that is comparable or smaller than the size of a focal-plane pixel ($\sim$1\,cm$^2$). By integrating an entire spectrometer into the footprint of a focal-plane pixel, it becomes possible to create arbitrarily large, densely-packed, and scalable IFU spectrometer arrays.}
\item{Direct detectors with NEP$\sim$3$\cdot$10$^{-18}$\,W/Hz$^{1/2}$, with footprint $\le$1\,mm$^2$ to allow packing $\sim$200 (or more) detectors onto a focal-plane pixel.}
\item{Multiplexed readout for order $\sim$20,000 mm-wave detectors in total.}
\end{enumerate}

Recent advances in technology have readily provided us with all three prerequisites, separately. On-chip mm-wave spectrometers, like SuperSpec\cite{SuperSpec-concept,SuperSpec}, DESHIMA\cite{DESHIMA, DESHIMA2}, or SPT-SLIM\cite{SPT-SLIM}, with $R$ of a few hundred have been demonstrated (albeit in considerably larger footprints than what we aim for). Crucially, the channelizer part of these is already much smaller than the typical $\sim$1\,cm$^2$ of a mm-wave focal-plane pixel of an envisaged mm-wave instrument (most of the real-estate taken up by the readout resonators of the KIDs for both). Highly sensitive KIDs that far surpass the NEP requirement for background-limited operation have been produced\cite{SPACEKIDS}, and KIDs that meet the size and multiplexing requirements are presently being deployed\cite{AMKID, AMKID-new} (albeit not as antenna-coupled spectrometer devices). What remains is combining these advances in the form of a new mm-wave integral field unit spectrometer that can fill the field-of-view (FoV) of a 10-m class or larger mm-wave telescope with densely packed spectrometer pixels. It could be done 'today'. This view is reinforced by the current development of TIFUUN\cite{TIFUUN} for ASTE\cite{ASTE} to realize a very similar IFU spectrometer to the one we propose here.

\section{Instrument Concept}
\label{sec:concept}

Spectroscopic galaxy surveys at mm-wavelengths could be conducted from ground-based facilities. Our posit is that a 100-pixel IFU spectrometer with $R$$\sim$200 could fill a 10$^\prime$ field-of-view of a 10-m class mm-wave facility, covering the entire 1.2\,mm atmopsheric window, and would detect up to 20,000 redshift complete star-forming galaxies down to near $L_*$ luminosities in 1000\,h of operation.

The 1.2\,mm atmospheric window is especially promising in terms of yielding the highest detection rates, through a combination of factors: (1) the widest fractional bandwidth observable from the ground ($\Delta f / \langle f \rangle$$\sim$0.5), which also results in the (2) lowest $z_{\rm min}$ cutoff, (3) low atmospheric opacities most of the time even at moderate altitudes, and (4) allows for cm$^2$-scale packed pixel footprints, and (5) a reasonably flat CO redshift selection for $z$$\sim$1.5--12, being firmly on the Rayleigh-Jeans side of the SED even for sources of extreme redshift ($z$$>$6). However, we'll consider designs for the neighboring 2\,mm and 850\,$\mu$m atmospheric windows also. See Tables~\ref{tab:bands}--\ref{tab:performance} for details.

We initially target $R$$\sim$200 spectral resolution for practical reasons. While the most optimal detection of typical extragalactic CO linewidths (which we guess to be in the range of $\Delta v$$\sim$300--600\,km/s on average) may push for $R$ in the 500--1000 range, it would require packing 2.5--5 times as many channels on a focal-plane pixel than an $R$$\sim$200 design. Thus, while an $R$$\sim$200 resolution may be somewhat less optimal for the detection of star-forming galaxies on average, it is significantly more feasible technologically, at least for now.

\begin{table}[htb]
\caption{$R$=200 spectrometer pixel properties.}
\centering
\begin{tabular}{|c|c|c|c|c|c|}
\hline
 $\bm{\lambda}$      &  {\bf Band}      & {\bf Bandwidth} & $\bm{z_{\rm min}}$ & {\bf octaves} & {\bf channels} \\
 (mm)           & (GHz)      & (GHz) & & & (count) \\ 
\hline
2     &  125 -- 175 &   50  & 0.31 &    0.5   & 200 \\
\hline
1.2   &  190 -- 310 &  120  & 0.15 &   0.7   & 280 \\
\hline
0.87  &  330 -- 360 &   30  & 2.9  &   0.13  & 52 \\
\hline
\end{tabular}
\label{tab:bands}
\end{table}

We baseline the IFU spectrometer concept for a 10-m class telescope. These are the smallest of the 'large' mm-wave telescopes, which abound thanks to many 12-m ALMA prototypes (e.g.\ APEX\cite{APEX}, GLT\cite{GLT}), as well as existing telescopes of other designs (e.g., LCT\cite{LCT} or SPT\cite{SPT}, ASTE\cite{ASTE}). The point is to show that a 10-m class telescope is perfectly capable delivering the transformative science when equipped with a background-limited $R$$\sim$200 IFU spectrometer, and operated from a site where PVW$\sim$2\,mm is typical.

We will assume a 10 arcmin FoV, which is achievable for most 10-m class mm-wave telescopes (e.g.\ in the Cassegrain cabin of the ALMA prototypes) without major modifications.

For the concept, we will assume $f/D$=5 optical configuration, which is not atypical for existing mm-wave cameras. The $f/5$ is used merely to estimate the physical size of 'pixels' in the focal plane. Accordingly, the 10 arcmin field-of-view translates to around 12\,cm on the focal plane, which is just below the size of a $\sim$5-inch wafer that is typically used for lithography. For larger focal planes, it may be necessary to tile multiple wafers to provide the desired total field of view. The concept is easily adjusted for different optical configurations also, and larger $F$ numbers are can accommodate larger physical pixels, hence provide more room for the integrated spectrometers. 

\begin{table}[hbt]
\caption{Estimated $R$=200 channel performance at PWV$\sim$2mm.}
\centering
\begin{tabular}{|c|c|c|c|}
\hline
 $\bm{\lambda}$      & {\bf NEFD}                & {\bf CO confusion limit}     & {\bf Conf.\ exposure} \\
 (mm)           & (mJy\,s$^{1/2}$)    & (mJy)                 & (h) \\ 
\hline
2     &   86    &  0.55  &    6.8   \\
\hline
1.2   &  220    &  1.1   &    11    \\
\hline
0.87  &  450    &  1.9   &    16    \\
\hline
\end{tabular}
\label{tab:limits}
\end{table}

The in-pixel $R$$\sim$200 channelizing of the incident radiation necessitates an antenna-coupled design, whereby light is coupled into a lithographic transmission line filterbank. Because of the relatively narrow-band ($R$$\sim$200) flux measurement, we want to focus as much of the incident light from a point-source onto a single pixel to achieve the best sensitivities. The incident radiation may be focused on the antenna by a suitable feedhorn or a lens. The most efficient coupling to the telescope beam is likely achieved with feedhorns packed at a 2$F\lambda$ spacing. To maximize sensitivity, ideally we couple both orthogonal linear polarizations for a given channel into the same total-power detector. Such dual-polarization driving ought to be possible for kinetic inductance detectors (KIDs), where a single detector could absorb power from two separate channelizer structures originating from orthogonally polarized antenna inputs.

Given the above parameters (10-m telescope, 10 arcmin FoV, $\lambda$$\sim$1.2\,mm, $F$=5, 2$F\lambda$ pixel spacing), a suitable IFU spectrometer would have around 100 pixels, arranged in a hexagonal pattern at $2F\lambda$ spacing, fully populating the available field-of-view. Each pixel, with a physical area of $\sim$1.1\,cm$^2$, must fit an $R$$\sim$200 channelizer for $\sim$280 Nyquist sampled channels, and a matching number of detectors per pixel, or 25 thousand detectors in all. Like for SuperSpec\cite{SuperSpec}, the channelizer can be provided via thin-film microstrip transmission lines, on an area that is $\le$20\,$\mu$m wide and $R\lambda/\varepsilon_r$ long. For the 1.2\,mm band that translates to an area of $\le$0.5\,mm$^2$ assuming a suitable dielectric with $\varepsilon_r$$\ge$10. Accordingly, around 99\% of the focal-plane real-estate can be used by the 280 detectors per pixel, which sets the size of individual detectors to $\le$0.44\,mm$^2$ (or equivalently $\le$[0.66\,mm]$^2$). That detector area is not too different from that of the KIDs currently being deployed in the high-frequency array (HFA) of the APEX MKID camera\cite{AMKID, AMKID-new}.

\begin{table}[tb]
\caption{$R$=200 $f/5$ IFU spectrometer and galaxy survey performance overview.}
\centering
\begin{tabular}{|c|c|c|c|c|c|c|}
\hline
 $\bm{\lambda}$      & {\bf FWHM}       & {\bf pixels}    & {\bf detectors}      &  $\bm{A_{\rm det}}$   & {\bf Mapping Speed}  &  $\bm{ N_{\rm gals}}$ \\
 (mm)           & (arcsec)   & (count)   & (count)        &  (mm$^2$)        & (deg$^2$/year) & (count/year) \\ 
\hline
2     &   42    &  37   &   7,400  & 1.52  &  0.82  & 10,800 \\
\hline
1.2   &   25    &  91   &  25,480  & 0.44  &  0.45  & 23,300 \\
\hline
0.87  &   18    & 169   &   8,788  & 1.29  &  0.30  & 5,500 \\
\hline
\end{tabular}
\label{tab:performance}
\end{table}

Table \ref{tab:performance} summarizes the expected performance and scientific yield of a IFU spectrometer with 10\,arcmin FoV on a 10-m telescope for 1 year ($\sim$1000 hours on source) of operation for the deep galaxy survey. Table \ref{tab:LIM} does the same for the line intensity mapping use case, assuming a 10\,deg$^2$ survey area. For the galaxy survey we show the confusion-limited survey area (i.e., mapping speed) and the expected number ($N_{\rm gals}$) of individually detected, redshift-complete galaxies, per year of operation. For LIM we identify the line species, which are expected to dominate the power spectrum signal, while noting also that a multitude of lines from different redshifts will also contribute to the autocorrelation power measured in each band. Shown are the estimated noise per mode ($P_N$), the total Number of modes measured ($N_{\rm modes}$), and the lowest wavenumber nominally measured by the survey ($k_{\rm min}$), assuming no limitations from atmospheric fluctuations or other sources of 1/$f$ noise.

\begin{table}[tb]
\caption{A 10\,deg$^2$ LIM survey after one year (1000\,h) of operation.}
\centering
\begin{tabular}{|c|c|c|c|c|c|}
\hline
 $\bm{\lambda}$      & {\bf Line}       & $\bm{z}$    & $\bm{P_N}$      &  $\bm{N_{\rm modes}}$  &  $\bm{ k_{\rm min}}$ \\
 (mm)                &                  &             & ($\mu$K$^2$ Mpc$^3$ / $h^3$) &  ($\times$ 10$^6$)   &  ($h$ / Mpc) \\ 
\hline
2     &  CO(3-2)    &  1.3054   &  13,000 &  5.4  & 0.0399 \\
\hline
1.2   &  CO(4-3)    &  0.8443   &  24,000 &  2.7  & 0.0485 \\
\hline
0.87  &  CO(6-5)    &  1.0047   &  58,000 &  7.3  & 0.0482 \\
\hline
\end{tabular}
\label{tab:LIM}
\end{table}

\subsection{Challenges}

While there are no technological gaps that would prevent the realization of a IFU spectroscopic imaging survey instrument as described, a successful implementation is expected to encounter a number of practical challenges nevertheless:

\begin{itemize}

\item{Stray-light and out-of-band radiation. Our instrument concept is based on broad-band power detectors (such as KIDs), which are coupled to the incident radiation from a celestial source through the telescope and instrument optics, an antenna, and a network of lithographic transmission lines. However, the detectors may also respond to unintended sources of direct illumination. Because direct illumination may be broad band, in contrast to the $R$$\sim$200 channels coupled via the antenna and filterbank, they must be suppressed $\sim$200 times below of what would be sufficient for a broad-band continuum camera. To avoid an undesired loss of sensitivity, the utmost care must be taken to avoid stray light and/or blue-leak contamination of the detectors, within the instrument, and in the telescope optical environment in which the instrument will operate.}

\item{Standing wave patterns, both on pixel and before it, can severely disrupt the spectrometer performance, resulting in a highly uneven and/or unstable spectral response across the observed pixel bandwidths. Signals typically travel an $\sim$$R\lambda$ total path length along the channelizer structure of the pixel alone. To avoid undesired standing waves developing along that long 'cavity', it may be necessary to terminate the channelizer structure with a broad-band absorber to prevent reflection at its end, which could adversely affect the performance of the spectrometer chip. One must also be careful to avoid on-axis reflections along all optical elements of the telescope and the instrument also, each of which could result in a similar performance degradation also.}

\item{Phononic isolation of the detectors may be necessary, both to keep power in the detectors localized long enough for effective translation into electronic signals, and to prevent cross-talk among the thousands of detectors residing on the same substrate. Given the number of detectors per pixel and overall, the cross-detector leakage of non-thermal phonons within a pixel needs to be suppressed at the $\ge$30\,dB level, while across pixels it may need to be suppressed at the $\ge$50\,dB to avoid an increase in the effective background noise levels due to cross-contamination.}

\item{Magnetic shielding. KIDs (and other superconducting detectors) can be sensitive to magnetic fields, especially during the cooldown phase, and therefore effective magnetic shielding of the detectors in the focal plane may be critical for achieving the nominal detector performance in a real-world operation.}

\item{Cryogenics. Achieving the necessary detector NEPs for background limited operation from the ground may require operation near or below 100\,mK, although the non-thermal detection mechanism of KIDs may also allow operating them at a higher temperature, such as 260--280\,mK without a significant performance degradation. Depending on the telescope it is operated on, the IFU spectrometer may have to be installed in the Cassegrain focus, in which case the dewar will tilt with telescope elevation, and pose additional challenges to maintaining cryogenic performance at the range of typical tilt angles during observations.}

\item{Readout. The small pixel footprint may require readout multiplexing at higher frequencies (e.g., $\sim$10\,GHz) for KIDs, although the use of parallel-plate capacitors, instead of the more typical inter-digitated ones, can also shrink the size of resonators significantly\cite{PRIMA-KIDS}, provided these can be produced reliably and with sufficient targeting precision also. A higher frequency readout may involve additional up- and downconverters to interface with digital-to-analog waveform outputs and the digitization of signals at a lower baseband.}

\item{Operational challenges. While not inherent to the design or physical realization of the instrument per se, a common operational challenge for massively multiplexed KIDs is identifying which resonance belongs to which physical detector (i.e., spatial-spectral voxel). This may be addressed with suitable calibration hardware, such as wire-grid scanners in orthogonal directions\cite{AMKID-new}, and a sweeping LO for the frequency direction, together which can 'map' resonators to voxels quickly. The readout position of resonators can be characterized and modeled in the lab.}

\end{itemize}

None of the challenges listed above are show-stoppers. All of them can be (and have been) mitigated effectively, as have been demonstrated by the various on-chip spectrometer instruments and KID cameras already deployed or under development. However, failing to adequately address them all can lead to degraded instrument perfomance, and result in significantly (or vastly) reduced scientific yield. Thus, the optimal realization of such an instrument will likely require the expertise and contributions from multiple persons, groups, and/or institutions.

At the same time, a spectroscopic imager is exempt from some of the constraints that affect similar single-beam spectrometers, like SuperSpec. Notably, a multi-beam spectrometer does not require that all spectral channels are operational on any given pixel, or that channels are spaced at closely regular intervals in frequency, or that they provide closely uniform sensitivities. As the celestial source is scanned across many/all pixels during an observation\cite{Scanning}, spectral information can be reconstructed effectively and uniformly as long as any given spectral 'channel' is 'typically' present, with a 'typical' sensitivity, among the ensemble of pixels. As such, detector yields as low as 70--80\% and/or gaps in the frequency coverage of individual pixels are generally tolerable, while non-uniformities will average out in the mapping process.

\section{Scoping and Variants}
\label{sec:scoping}

There is not one way to realize the IFU spectrometer for the galaxy or LIM surveys. Depending on the availability of funding, resources, telescope, preference for particular technologies, and the degree of ambition, there is ample room for adjusting the concept for specific needs or constraints. For example, one may scope (or descope) specific aspects of the instrument concept, or utilize alternative technologies to what we considered so far.

\subsection{Telescope-specific variants}

The instrument can be (and should be) optimized for the telescope it is ultimately deployed on, such as:

\begin{itemize}
\item{Larger telescopes (e.g.\ JCMT, IRAM 30-m\cite{IRAM-30m}, LMT\cite{LMT}, or AtLAST\cite{AtLAST}) could do the same type of surveys faster and/or go deeper still. A 50-m class telescope may also provide astrometry of the sources at the 1--2$^{\prime\prime}$ level, which is sufficient for direct follow-up in most other bands.}

\item{The IFU spectrometer field-of-view may be scaled to match what the host telescope optics can deliver. As the survey speed scales with the both the instrument's FoV and with the telescope aperture size, it is generally desirable to fill the available FoV with pixels to the maximal extent possible, while larger telescopes, such as the IRAM 30-m, can deliver same results with a smaller FoV also.}

\subsection{Capabilities}

There is flexibility on tuning the capabilities that the IFU spectrometer is designed to deliver, e.g.:

\item{Any number of bands (1--4), and any mm-wave band (850\,$\mu$m -- 3\,mm) can be scientifically interesting and useful -- for deep CO galaxy surveys and LIM application alike. Note, however, that for a multi-band instrument many of the technological challenges mentioned become aggravated.}

\item{Spectral resolution can be interesting and useful anywhere between $R$$\sim$150 ($\Delta v$$\sim$2000\,km/s) and $R$$\sim$1000 ($\Delta v$$\sim$300\, km/s). Low $R$ values may suffer from the dilution of the CO signal, when the line widths of the star-forming gas in the targeted galaxies do not fill the spectral channels. High $R$ values may be challenging to realize, and may also result in loss sensitivity if the galactic lines spread over multiple spectral bins. Optimal sensitivities are reached when the channel widths closely match the typical (mean) projected velocity dispersion of the star-forming gas. Assuming that $\pm$250\,km/s projected velocity spreads are typical, $R$$\sim$600 may be more optimal from a sensitivity perspective alone, but other constraints may push toward targeting different $R$ values.}

\subsection{Technologies}

There are also technology driven choices that instrument builders can make. For example:

\item{In this paper we focused on in-plane, on-chip spectrometers, but there are alternatives to achieving compactness, such as waveguide spectrometers\cite{WSPEC} or vertical integration\cite{Stover2024}.}

\item{The pixel packing can be adjusted. The $2F\lambda$ spacing maximizes the survey speed for a fixed number of pixels\cite{Holland2002}, and relaxes the NEP requirement by focusing a full telescope beam on an individual pixel. However, smaller, more tightly packed pixels are also viable provided that the necessarily lower detector NEP values can be achieved for background-limited operation.}

\item{We envision a feedhorn-coupled design illuminating the antenna(s) on pixels, because horns provide a naturally effective way for rejecting stray and out-of-band light. However, lens coupling may be viable also, provided that the necessary degree of rejection of stray and out-of-band light can be achieved otherwise.}

\item{Dual-polarization sensitivity may be achieved via suitable dual-polarization antenna designs\cite{Bueno2018} also.}

\item{It is not necessary for the IFU spectrometer to have dual-polarization sensitivity. A single-polarization instrument can do the same survey science. It just needs twice the time to reach equivalent sensitivities.}

\item{The $F$ number may be chosen to any convenient value for the implementation, and suitable re-imaging optics can interface the instrument to the telescope, as appropriate.} 

\end{itemize}

\section{Implementation}
\label{sec:implementation}

We initially conceived of building such an IFU spectrometer for the Greenland Telescope (GLT) specifically\cite{GLT-survey}, since the Center for Astrophysics $|$ Harvard \& Smithsonian (CfA) is a partner institution of the GLT. Tim Norton was at the time the director of the GLT, while Thomas Greve is involved with the GLT from the Danish side. It has been clear to us from the outset that we need to collaborate with other groups and institutions to develop, build, and deploy such an instrument. It is also clear that other telescopes would be just as suitable platforms for the same type of instrument or science.

The proposed instrument is sufficiently complex that implementation within a single institution may prove challenging, as demonstrated by the most recent wave of large cameras with similar scope and complexity\cite{SCUBA-2, TolTEC, AMKID}. It is increasingly rare that a single institution would have all the resources (financial, technological, and human), as well as all the in-house expertise, necessary to cover every aspect of design, production, and fielding optimally; and to deliver flawless instrumentation on the typical timescales of funding cycles (3--5 years).

We therefore propose a new approach. It relies on the loose, open-source collaboration of partners, and benefits from the contribution of leading experts from all over the world. For example, participants who join might generally agree to the following foundational principles:

\begin{itemize}
\item{The collaboration is open to all who wish to contribute to it.}
\item{Participants agree to abide by a code of honor. (We are in this together because we all want to see the best IFU spectrometer(s) be built and operated based on the concepts and designs we develop together.)}
\item{All information shared with the collaboration is considered private knowledge, not to be re-shared outside of the collaboration without the explicit consent by those who contributed the information.}
\item{Any participant in the collaboration may freely use, adapt, modify, or innovate any aspect of the IFU spectrometer concept in any way they deem appropriate, as long as they re-share all relevant new information (e.g.\ designs, simulations, lab measurements, documentation, publications) with the the rest of collaboration.}
\item{Any participant is free to work independently, secure their own funding, or form their own working partnerships within the collaboration, for example to build and operate a particular implementation of the shared concepts. There is no need for approval or even coordination with other participants of the collaboration on their individual activities.}
\item{Any participant is free to publish or present their own work to the public, as they see fit, and with the list of authors they deem appropriate, as long as they simply acknowledge the collaboration, and appropriately cite or acknowledge the contributions of others.}
\end{itemize}

These principles are meant to protect innovation while fostering a free-spirited collaboration among the partners. All information shared within the collaboration will likely be through a properly authenticated version control system (e.g.\ GitHub\footnote{\url{https://www.github.com}}), such that once information is shared it remains accessible to all involved, but not to outside the collaboration. (The details of the version control system, organization, and workflows are to be identified later, and may evolve with the project.) Apart from sharing ideas, designs, and measurements, the collaboration will support its participants in any way possible to help them achieve their goals.

The above listed 'rules' of the collaboration are a starting point only. Ultimately, a collaboration, once formed, may decide a different set of rules, which promote progress and protect innovation in the way the participants feel is best for them.

\subsection{Modular Designs}

While there is a uniting concept centered around a IFU spectrometer capable of delivering the next-generation cosmological surveys of star-formation, it is not limited to the realization of a singular instrument or design that alone is capable of delivering on that goal. For every design choice one can make, alternative choices may be viable also. As such, our aim is to compile a small library of designs and components from which an instrument builder may draw to assemble a particular implementation of the IFU spectrometer concept. We may, therefore, have multiple alternatives for the detectors, focal-plane units, focusing elements, etc. And each module may in turn have sub-versions optimized somewhat differently, offering varying trade-offs. (Hopefully, common sense will guide us to converge on no more than a few alternative designs for the critical components).

As we progress, we may want to agree on certain specific standards (such as physical dimensions, or other specifications) for the components to allow swapping one design with another easily, e.g.\ to accommodate changing requirements or budgetary constraints, or to support variant implementations. As the collaboration grows and evolves, we shall identify areas where such standardization is desirable or necessary, in order to facilitate continued progress.

\section{Conclusion}

We envision that a mm-wave IFU spectrometer can be designed, developed, and built on a $\sim$5-year timescale, and deployed to a suitable 10-m class (or larger) ground-based mm-wave telescope to conduct the first truly volumetric surveys of star-formation in the Universe, in a dedicated effort over 1--3\,years of operation. Such IFU spectrometers can for the first time deliver a complete census of star-formation at $z$$\sim$1--12, individually detecting tens of thousands of star-forming galaxies above a mass-luminosity threshold, complete with spectroscopic redshifts for each and every one. These types of instruments will be able to probe an order of magnitude deeper in star-forming galaxy masses and luminosities than any previous degree-scale mm-wave survey to date, and for the first time approach the $L_{*}$ galaxy populations that represent that bulk of the galaxies in the Universe. The same or similar instrument can conduct Line Intensity Mapping (LIM) also, and the two complementary surveys can be conducted in a time-sharing arrangement for maximum scientific impact.

We believe that such an IFU spectrometer, or family of instruments, are best built through a loose, open-source collaboration that involves a wide range of experts from around the globe, and in which participants are free to pursue their own specific goals and directions without being confined by the restrictive rules that are typical to consortia. 

We hope that you, the reader of this paper, will join this collaboration to contribute your expertise to make these IFU spectrometers the best they can be. If you are an instrument builder, we hope that you join because you can draw on the knowledge shared in the collaboration to build your capable instrument, or component, to its utmost potential and without delays and detours. And, we hope telescope organizations will join because they want a share of the transformative science that these IFU spectrometers will deliver when deployed on a telescope they run.

It you have questions about the collaboration, or wish to be a part of it, please contact us by email at the address provided on the title page of this article.

\subsection* {Disclosures}

The authors declare that there are no financial interests, commercial affiliations, or other potential conflicts of interest that could have influenced the objectivity of this research or the writing of this paper.

\subsection* {Code, Data, and Materials Availability} 

All data in support of the findings of this paper are available within the article or as supplementary material.

\subsection*{Acknowledgements}

AK, GK, and TN acknowledge support from the the Smithsonian Astrophysical Observatory.

TRG acknowledges financial support from the Cosmic Dawn Center (DAWN), funded by the Danish National Research Foundation (DNRF) under grant No. 140. TRG is grateful for support from the Carlsberg Foundation via grant No. CF20- 0534.

A previous version of this paper was published in the Proceedings of the SPIE Conference 13101\cite{spie-paper}.

\bibliography{survey} 
\bibliographystyle{spiejour} 

\vspace{2ex}\noindent\textbf{Attila Kov\'{a}cs} is a computer engineer at the Center for Astrophysics $|$ Harvard \& Smithsonian. He received his A.B. degree in physics, astronomy and astrophysics from Harvard College in 1997, and his PhD degree in physics from the California Institute of Technology in 2006. His current research interests include submillimeter detector technologies and instrumentation, star-forming galaxies, signal processing, and data reduction and imaging algorithms.

\vspace{2ex}\noindent\textbf{Garrett K. Keating} is an astrophysicist at the Center for Astrophysics $|$ Harvard \& Smithsonian and is the Deputy Director of the Submillimeter Array. He received his B.A. in astrophysics from the University of California, Berkeley in 2008, and his Ph.D. in astrophysics from the University of California, Berkeley in 2016. His current research interests includes radio instrumentation, molecular gas in the early Universe, supermassive black holes, and atmospheric effects on interferometric observations.

\vspace{2ex}\noindent\textbf{Thomas R. Greve} is a professor and co-director of the Cosmic Dawn Center, a center of excellence at the Technical University of Denmark and Copenhagen University since 2018. From 2012, Associate Professor at University College London. Obtained his PhD in 2005 from the Institute for Astronomy Edinburgh, followed by postdoctoral research positions at the California Institiute of Technology and the Max-Planck Institute for Astronomy. He led some of the first systematic surveys of molecular gas and dust in extremely star-forming galaxies in the young Universe. Since 2020, the Danish co-PI on the MIRI instrument on JWST, and the MIRI Deep Imaging Survey (MIDIS) of the Hubble Ultra Deep Field.

\vspace{2ex}\noindent\textbf{Timothy Norton} is Director of the Submillimeter Array, Hawaii.

\vspace{1ex}
\noindent Biographies and photographs of the other authors are not available.

\listoffigures
\listoftables

\end{spacing}
\end{document}